\documentclass[prl,amsmath,amssymb,twocolumn,superscriptaddress,aps,floatfix]{revtex4-2}
\usepackage{subfigure}
\usepackage{graphicx}
\usepackage{dcolumn}
\usepackage{bm}
\usepackage{standalone}
\begin{document}

\preprint{APS/123-QED}

\title{ Dzyaloshinskii-Moriya Induced Topological Magnon-Phonon Hybridization in 2D Antiferromagnetic Insulators with Tunable Chern Numbers}

\author{Bowen Ma}
\affiliation{Department of Physics, The University of Texas at Austin, Austin, Texas 78712, USA\\
}

\author{Gregory A. Fiete}
\affiliation{Department of Physics, Northeastern University, Boston, Massachusetts 02115, USA\\
}
\affiliation{Department of Physics, Massachusetts Institute of Technology, Cambridge, Massachusetts 02139, USA}

\date{\today}

\begin{abstract}
We theoretically study magnon-phonon hybrid excitations (magnon-polarons) in two-dimensional antiferromagnets on a honeycomb lattice. With an in-plane Dzyaloshinskii-Moriya interaction (DMI) allowed from mirror symmetry breaking from phonons, we find non-trivial Berry curvature around the anti-crossing rings among magnon and both optical and acoustic phonon bands, which gives rise to finite Chern numbers. 
We show that the Chern numbers of the magnon-polaron bands can be manipulated by changing the magnetic field direction or strength. We evaluate the thermal Hall conductivity reflecting the non-trivial Berry curvatures of magnon-polarons and propose a valley Hall effect resulting from spin-induced chiral phonons as a possible experimental signature. Our study complements prior work on magnon-phonon hybridized systems without optical phonons and suggests possible applications in spin caloritronics with topological magnons and chiral phonons.
\end{abstract}

\maketitle

\paragraph{Introduction.}Antiferromagnetic materials have recently attracted a great deal of attention within the community of spintronics \cite{jungwirth2016antiferromagnetic,jungwirth2018multiple,baltz2018antiferromagnetic}, because they are rather insensitive to the perturbation of magnetic fields and have small stray fields with fast THz magnetic dynamics compared to ferromagnets with frequencies in the GHz range.  Research over the past decade has focused on spin dynamics and spin transport in antiferromagnets, which may originate from spin-transfer torques \cite{cheng2014spin,gomonay2010spin}, domain-wall motion \cite{gomonay2016staggering}, and the spin Seebeck effect \cite{wu2016antiferromagnetic,ohnuma2013spin,rezende2016theory}. Magnons, as collective excitations emerging from magnetic order, have low-dissipation and permit a pure spin transport without Joule heating, leading to a surge of interest in utilizing magnons for spintronics. Many magnonic analogs of electronic phenomena, such as the magnon thermal Hall effect \cite{katsura2010theory,onose2010observation,matsumoto2011rotational}, the magnon Nernst effect \cite{cheng2016spin,zyuzin2016magnon,shiomi2017experimental} and the magnonic Edelstein effect \cite{li2020magnonic,zhang2020magnon}, have been theoretically studied and experimentally observed.

Along with magnonics, there is also a potential application in spintronics by combining magnetic orders with non-trivial band topology \cite{vsmejkal2018topological}. Topologically protected states are usually robust and only weakly affected by disorders. They can provide a high charge-to-spin conversion efficiency \cite{wang2016surface}, exhibit strong magnetoresistance \cite{wang2012room,liang2015ultrahigh} and possess a number of exotic phenomena such as the quantum anomalous Hall effect \cite{chang2013experimental,deng2020quantum} and chiral Majorana fermions \cite{he2017chiral}. In addition to fermionic topological excitations, there is also an emerging field of investigating topological bosonic excitations, such as topological magnons \cite{zhang2013topological,mook2014edge,chisnell2015topological,li2016weyl} and topological phonons \cite{prodan2009topological,zhang2010topological,jin2018recipe}. Moreover, some recent works have shown topological properties in hybridized systems between magnons and acoustic phonons with the magnetoelastic coupling \cite{park2019topological,go2019topological,zhang20203}, the Dzyaloshinskii-Moriya interaction (DMI) \cite{zhang2019thermal,park2020thermal}, and the dipolar coupling \cite{takahashi2016berry}. However, a study of the coupling between magnons and optical phonons is still lacking.

In this Letter, we study hybrid magnon-phonon excitations in a 2D collinear antiferromagnetic insulator (AFI) on the honeycomb lattice. The topological magnon bands originate from an in-plane nearest-neighbor DMI permitted by mirror symmetry breaking \cite{di2015direct,tacchi2017interfacial,qaiumzadeh2018theory}, which can be generically achieved in 2D van der Waals heterostructures, in the presence of magnon-phonon coupling. Since van der Waals antiferromagnets naturally possess at least two sublattices, it is possible to realize the coupling between magnons and optical phonons. In such a coupled magnon-(optical) phonon system, which has not been studied in the ferromagnetic case \cite{zhang2019thermal}, we find finite Berry curvature and non-zero Chern numbers. 

We also show that the Chern numbers of magnon-polaron bands can be manipulated by an external magnetic field. For connection to experiments, we evaluate the thermal Hall conductivity and propose a spin-induced valley Hall effect as a possible experimental observation. We emphasize that our results are generic to many lattice structures and can be easily generalized to three-dimensional systems, as discussed at the end of this Letter. Our work suggests antiferromagnets with multiple sublattices--in contrast to ferromagnets--serve as promising platforms to realize tunable topological excitations hybridizing magnons with both acoustic and optical phonons, where the topology of the bands can provide robust information transport and may find possible applications in spintronics.

\begin{figure}
\subfigure[]{\label{fig:Lattice}\includegraphics[width=5.8 cm]{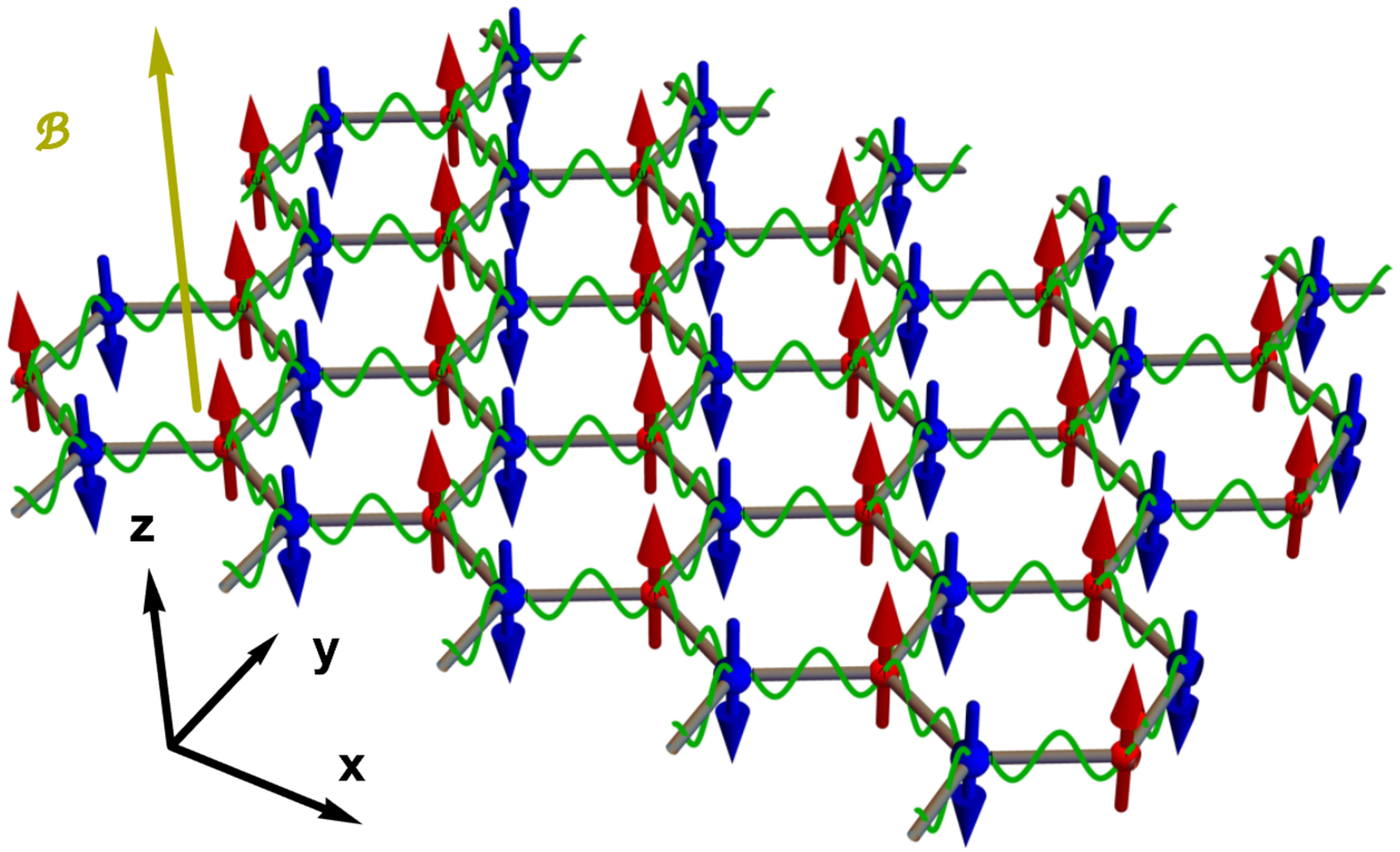}}\subfigure[]{\label{fig:DMV}\includegraphics[width=2.2 cm]{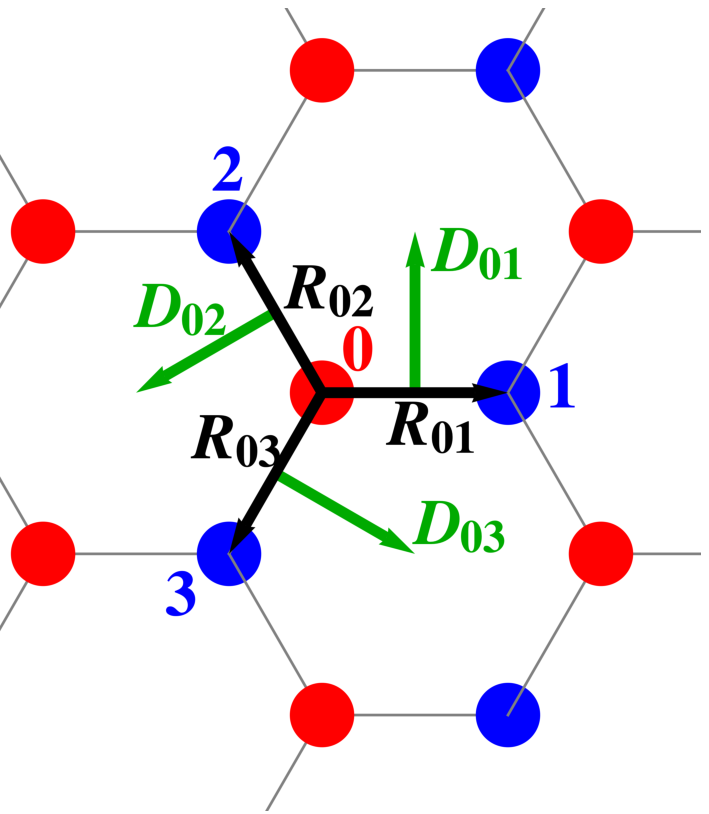}}
    \caption{(Color online.) (a) Schematic illustration of a hybrid magnon-phonon system. The ground state of the magnetization is Neel order along the $z$-axis (red and blue arrows, color denoting the A and B sublattices). (b) DM vectors (green arrows) for the nearest bonds originated from mirror symmetry $\mathcal{M}_{yz}$ breaking.}
    \label{fig:Schematic}
\end{figure}

\paragraph{Model.}We consider a system with collinear AFI Neel order on a honeycomb lattice, where the magnetic moments are perpendicular to the plane, i.e., $\mathbf{S}_{A,B}=\pm S\hat{\mathbf{z}}$ for the A and B sublattices respectively [see Fig.~\ref{fig:Lattice}]. The Hamiltonian describing both spin and lattice degree of freedom can be written as $H=H_m+H_p+H_{mp}$, where the magnetic part $H_m$ is given by,
\begin{align}
    H_m&=J_1\sum_{\left<ij\right>}\mathbf{S}_i\cdot\mathbf{S}_j-J_2\sum_{\left<\left<ij\right>\right>}\mathbf{S}_i\cdot\mathbf{S}_j\nonumber\\
    &-\frac{K_z}{2}\sum_i(S_i^z)^2-\mathcal{B}\sum_iS_i^z,\label{Hm}
\end{align}
where $J_1\ (J_2)>0$ is the (next-)nearest-neighbor antiferromagnetic (ferromagnetic) Heisenberg exchange coupling, $K_z>0$ is the easy-axis anisotropy and $\mathcal{B}=g\mu_BB$ is the external effective Zeeman magnetic field. The phonon part $H_p$ can be expressed as,
\begin{equation}
    H_p=\sum_{i}\frac{\mathbf{p}_i^2}{2M_i}+\frac{k_1}{2}\sum_{\left<ij\right>}(\hat{\mathbf{R}}_{ij}^0\cdot\mathbf{u}_{ij})^2
    +\frac{k_2}{2}\sum_{\left<\left<ij\right>\right>}(\hat{\mathbf{R}}^0_{ij}\cdot\mathbf{u}_{ij})^2,\label{Hp}
\end{equation}
where $\mathbf{u}_{ij}=\mathbf{u}_{j}-\mathbf{u}_{i}$ is the in-plane displacement of the lattice, $\hat{\mathbf{R}}_{ij}^0$ is the unit vector along bond $ij$ in equilibrium, and $k_1\;(k_2)$ is the spring constant that corresponds to the elastic energy between two (next) nearest neighbor ions. Here we ignore out-of-plane vibrations as they are higher-order terms \cite{sm}.

For the magnon-phonon coupling $H_{mp}$, we begin from an in-plane nearest-neighbor DMI originating from mirror symmetry breaking. By Moriya's rule \cite{moriya1960anisotropic,moriya1960new}, the direction of the DM vectors is perpendicular to the bond, i.e., $\mathbf{D}_{ij}\propto\hat{\mathbf{z}}\times\mathbf{R}_{ij}$ [see Fig.~\ref{fig:DMV}]. The DMI Hamiltonian is then
\begin{align}
    H_D=\mathbf{D}_{ij}\cdot(\mathbf{S}_i\times\mathbf{S}_j).\label{HD}
\end{align}
This term is not included in Eq.~(\ref{Hm}) since it is well-known that DM vectors perpendicular to spin moments do not appear in the linear spin-wave Hamiltonian \cite{katsura2010theory,zhang2019thermal,park2020thermal} and we assume it does not appreciably change (i.e., the change is numerically small) the Neel ground state order as long as the exchange coupling and anisotropy is large enough. However, both the magnitude and direction of $\mathbf{D}_{ij}$ depend on $\mathbf{R}_{ij}$ and thus it couples lattice and spin degrees of freedom. To lowest order, Eq.~(\ref{HD}) can be expanded as \cite{sm},
\begin{align}
    H_{mp}&\approx\frac{DS}{a}\sum_{\left<ij\right>}\mathbf{u}_{ij}\left[\mathcal{I}_{2}-\hat{\mathbf{R}}_{ij}^0\hat{\mathbf{R}}_{ij}^0\right]\left(\tilde{\mathbf{S}}_{A,i}+\tilde{\mathbf{S}}_{B,j}\right)\nonumber\\
    &=\frac{DS}{a}\sum_{\left<ij\right>}\left(\hat{\mathbf{R}}_{ij}^0\times \mathbf{u}_{ij}\right)\cdot\left[\hat{\mathbf{R}}_{ij}^0\times\left(\mathbf{S}_{A,i}+\mathbf{S}_{B,j}\right)\right],\label{Hmp}
\end{align}
where $D=|\mathbf{D}_{ij}|$ is the magnitude of the DMI, $a=|\mathbf{R}^0_{ij}|$ is the bond length, $\mathcal{I}_{2}$ is the $2\times2$ identity matrix, $\hat{\mathbf{R}}_{ij}^0\hat{\mathbf{R}}_{ij}^0$ is the Kronecker product between two $\hat{\mathbf{R}}_{ij}^0$'s and $\tilde{\mathbf{S}}_{A(B),i}=(S_{A(B),i}^x,\ S_{A(B),i}^y)$. The second equation mimics a Rashba-type spin-orbital coupling \cite{bychkov1984properties,manchon2015new} or a Raman spin-phonon interaction \cite{zhang2010topological,sheng2006theory,kagan2008anomalous,wang2009phonon}, which has been studied in topological aspects of spin or phonon systems. 

It is clear from Eq.~(\ref{Hmp}) that the DMI induced magnon-phonon coupling breaks the combined symmetry of time reversal plus $180^{\circ}$ rotation 
about an in-plane axis \cite{chen2014anomalous,suzuki2017cluster}. With magnetic fields, this symmetry breaking allows the existence of a thermal Hall effect \cite{mook2019thermal}, which is absent in a magnon-only or phonon-only scenario. Moreover, in contrast to the ferromagnetic case, $H_{mp}+H_m$ also breaks inversion symmetry \cite{cheng2016spin} and gives rise to chiral phonons at high symmetry points \cite{zhang2014angular,zhang2015chiral}, as will be shown below.

\begin{figure}
\subfigure[$\mathcal{B}=0.3$ meV]{\label{fig:LH}\includegraphics[width=4.3 cm]{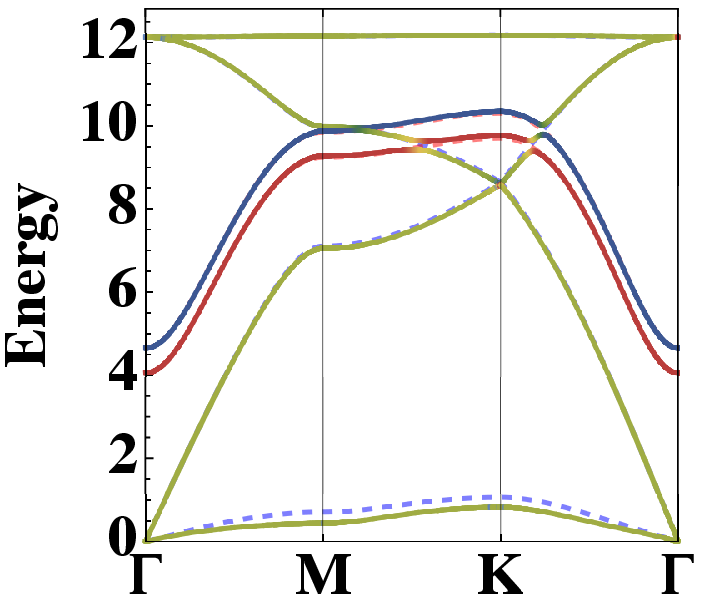}}\subfigure[$\mathcal{B}=0.6$ meV]{\label{fig:HH}\includegraphics[width=4.3 cm]{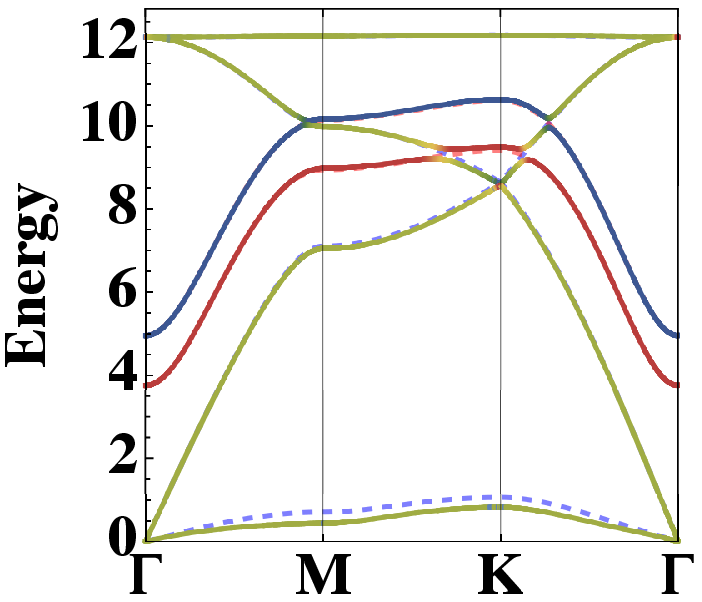}}
\subfigure{\includegraphics[width=4.3 cm]{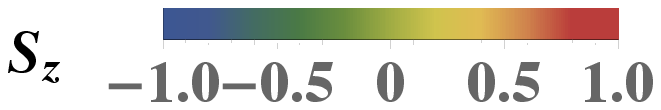}}\setcounter{subfigure}{2}
\subfigure[$\mathcal{B}=0.3$ meV]{\label{fig:LH_Detail}\includegraphics[width=4.3 cm]{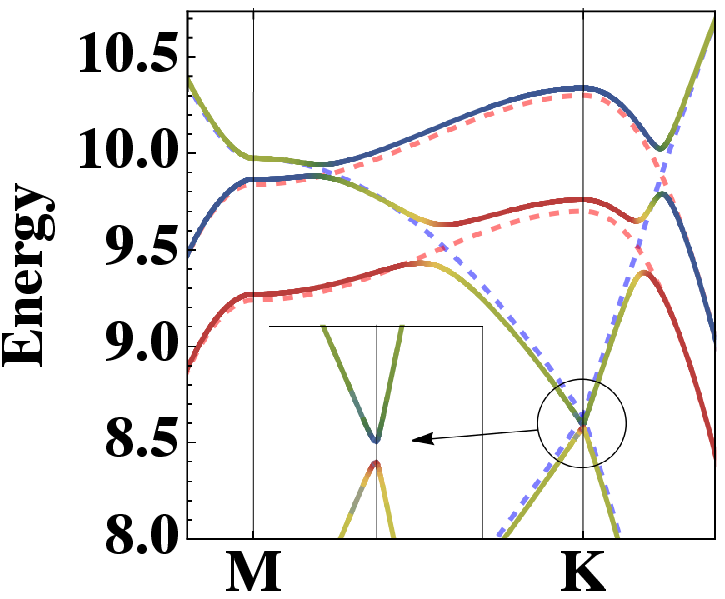}}\subfigure[$\mathcal{B}=0.6$ meV]{\label{fig:HH_Detail}\includegraphics[width=4.3 cm]{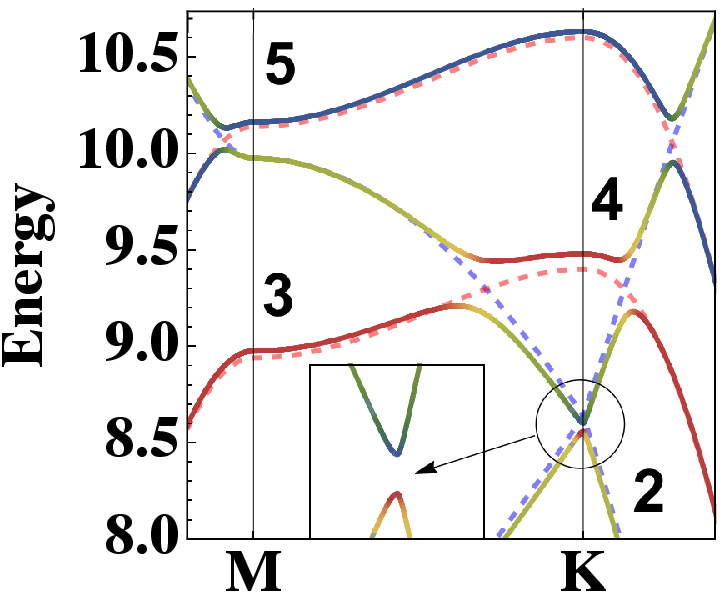}}
\subfigure[$\mathcal{B}=0.3$ meV]{\label{fig:CP_LH+}\includegraphics[width=4.3 cm]{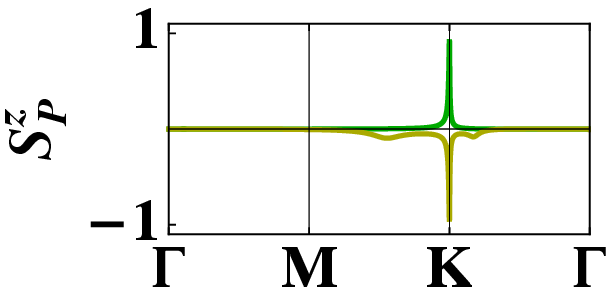}}\subfigure[$\mathcal{B}=0.6$ meV]{\label{fig:CP_HH+}\includegraphics[width=4.3 cm]{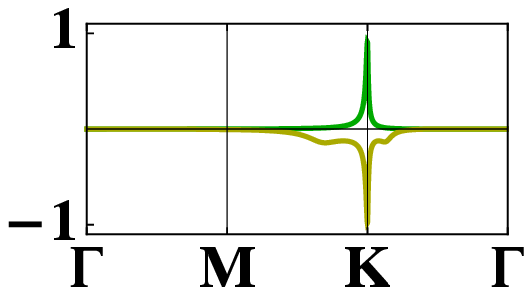}}
\subfigure[$\mathcal{B}=-0.3$ meV]{\label{fig:CP_LH-}\includegraphics[width=4.3 cm]{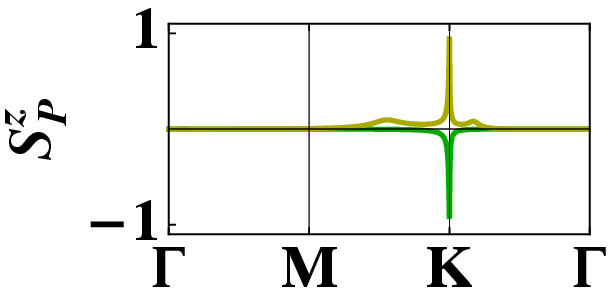}}\subfigure[$\mathcal{B}=-0.6$ meV]{\label{fig:CP_HH-}\includegraphics[width=4.3 cm]{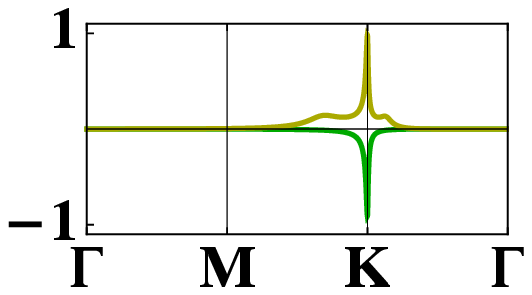}}
    \caption{(Color online.) Topological magnon-polaron bands. Energy is in meV. We set parameters as $S=3/2$, $J_1=2.0$ meV, $J_2=0.0$ meV, $K_z=1.0$ meV, $m_B/m_A=1$, $\hbar\sqrt{k_1/m_A}=7.0$ meV, $\hbar\sqrt{k_2/m_A}=0.5$ meV, $D=0.2$ meV.
The solid lines are bands with DMI and the blue (red) dashed lines are phonon (magnon) dispersions without DMI. (a)(b) Full band dispersions along high symmetry path. (c)(d) Bands around anti-crossing points. Band numbering is shown in (d). The insets show the gap opens at ${\bf K}$ and allows phonons with different chiralities (red or blue). (e-h) Phonon polarization contribution to $S_z$. The green (yellow) line is for band 2 (3).}
    \label{fig:Dispersions}
\end{figure}

\paragraph{Band Topology.}As magnons and phonons are both bosons, one can treat them equivalently as magnon-polaron excitations and re-write $H=H_m+H_p+H_{mp}$ to a generalized BdG form as \cite{sm},
\begin{align}
    H_\mathbf{k}=\left[
    \begin{array}{ccc}
        \frac{1}{2}\tilde{H}_{m}(\mathbf{k}) & \tilde{H}_{mp}(\mathbf{k}) & 0 \\
         \tilde{H}_{mp}^\dag(\mathbf{k})& \frac{1}{2}D(\mathbf{k})&0\\ 0&0&\frac{\mathcal{I}_{4}}{2M} 
    \end{array}\right],\label{Hk}
\end{align}
with representation $\mathbf{X}_\mathbf{k}=\left(a_{\mathbf{k}},b_{\mathbf{k}},a^\dag_{-\mathbf{k}},b^\dag_{-\mathbf{k}},\mathbf{u}_\mathbf{k},\mathbf{p}_{-\mathbf{k}}\right)^T$, where $a_{\mathbf{k}}\;(b_{\mathbf{k}})$ is the A (B) sublattice magnon annihilation operator in a Holstein-Primakoff representation \cite{holstein1940field}, $S^+_A\;(S^+_B)=\sqrt{2S}a\;(b^\dag)$, $\mathbf{u}_\mathbf{k}\ (\mathbf{p}_{-\mathbf{k}})$ is a four-vector for two-dimensional displacements (momenta) of A and B sublattices, $\tilde{H}_{m}\;(\mathbf{k})\ (\tilde{H}_{mp}(\mathbf{k}))$ corresponds to Eq.~(\ref{Hm}) [Eq.~(\ref{Hmp})] and $D(\mathbf{k})$ is the dynamical matrix corresponding to Eq.~(\ref{Hp}). Under this representation, the bosonic commutator is written as
\begin{align}
    \left[\mathbf{X}_\mathbf{k},\ \mathbf{X}_\mathbf{k}^\dag\right]=g=\left[\begin{array}{cccc}
        \mathcal{I}_{2}& &  &\\
        &-\mathcal{I}_{2} & & \\
        & & &i\mathcal{I}_{4}\\
        & &-i\mathcal{I}_{4} &
    \end{array}\right],\label{g}
\end{align}
and the eigenstates satisfy \cite{del2004quantum,shindou2013topological}, 
\begin{align}
 gH_\mathbf{k}\left|\psi_{n\mathbf{k}}\right>=\sigma_{nn}E_{n\mathbf{k}}\left|\psi_{n\mathbf{k}}\right>,\ 
 \left<\psi_{n\mathbf{k}}\right|g\left|\psi_{n'\mathbf{k}}\right>=\sigma_{nn'},
\end{align}
where $\sigma=\sigma_z\bigotimes\mathcal{I}_{6\times 6}$ stands for particle-hole space. With particle-hole symmetry, $E_{n\mathbf{k}}=E_{n+6,-\mathbf{k}}$ and thus we only plot the first six eigenvalues in Fig.~\ref{fig:Dispersions} and others are redundant. Here $S^z=\left<\psi^R_{n\mathbf{k}}\right|\left(-a_\mathbf{k}^\dag a_\mathbf{k}+b_\mathbf{k}^\dag b_\mathbf{k}+\mathbf{u}^A_\mathbf{k}\times\mathbf{p}^A_{-\mathbf{k}}+\mathbf{u}^B_\mathbf{k}\times\mathbf{p}^B_{-\mathbf{k}}\right)\left|\psi^R_{n\mathbf{k}}\right>$ mediates both magnon spins and phonon polarizations \cite{zhang2014angular}.

In Fig.~\ref{fig:Dispersions}, there are gapped rings around $\mathbf{\Gamma}$ or $\mathbf{K}\;(\mathbf{K}')$ formed by anti-crossing points among magnon and phonon bands due to the DMI coupling, which gives rise to nontrivial topological properties in this magnon-polaron system. In such a generalized BdG system, the Berry curvature is given by the Bloch wavefunction $\left|u_{n\mathbf{k}}\right>=e^{-i\mathbf{k}\cdot\mathbf{r}}\left|\psi_{n\mathbf{k}}\right>$ as \cite{matsumoto2011rotational,sm}
$\mathbf{\Omega}_{n\mathbf{k}}=i\left<\nabla_\mathbf{k}u_{n\mathbf{k}}\right|g\times\left|\nabla_\mathbf{k}u_{n\mathbf{k}}\right>,$
and the Chern numbers can be obtained by integrating Berry curvature $\Omega_{n\mathbf{k}}^z$ along the Brillouin zone as \cite{qi2008topological}
$
 C_n=\frac{1}{2\pi}\int_{BZ}d^2\mathbf{k}\;\Omega_{n\mathbf{k}}^z, 
$
from which we calculate the band Chern numbers [since the top (also bottom) two bands are degenerate at $\Gamma$ point, we add up the Berry curvature of the two bands to obtain a well-defined Chern number] by the Fukui method \cite{sm,fukui2005chern} and find that the magnetic field can change the Chern numbers by integers. 

In Fig.~\ref{fig:LH}, the Chern numbers for the middle three anti-crossed bands from low to high are $(-2, +4, -2)$, while they change to $(-2, +1, +1)$ in Fig.~\ref{fig:HH} by a phase transition when $\mathcal{B}>\mathcal{B}_c(\approx0.41$ meV with the parameters in Fig.~\ref{fig:Dispersions} \cite{sm}). Since in this parameter region the coupling barely affects acoustic modes and the top optical mode, the band topology can be effectively mapped into an SU(3) algebra \cite{barnett20123,zhang20203}. Here, instead of an analytic calculation (which is generally not accessible), we achieve an understanding of the band topology more intuitively by looking at Berry curvatures.

\begin{figure}
\subfigure[$\mathcal{B}=0.3$ meV]{\includegraphics[width=8.6 cm]{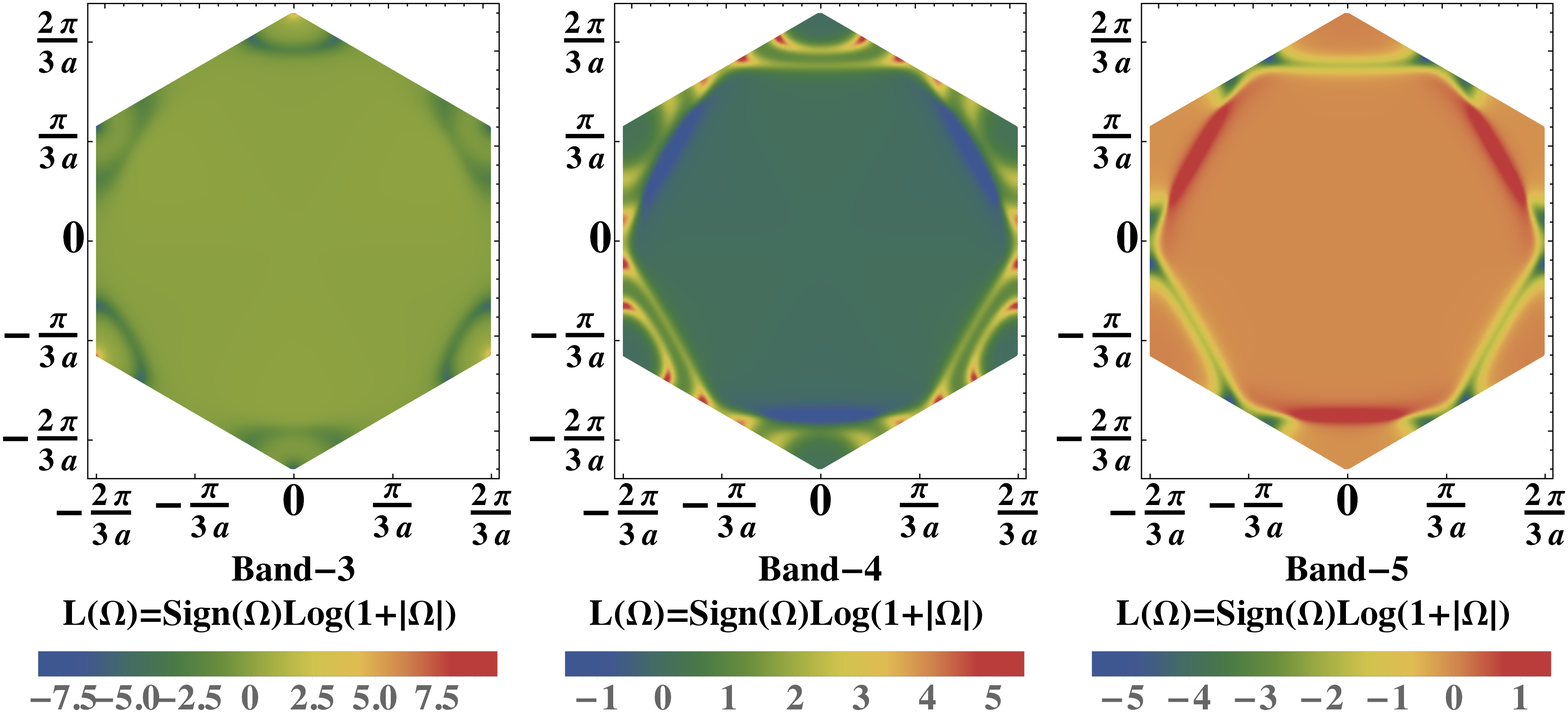}}
\subfigure[$\mathcal{B}=0.6$ meV]{\includegraphics[width=8.6 cm]{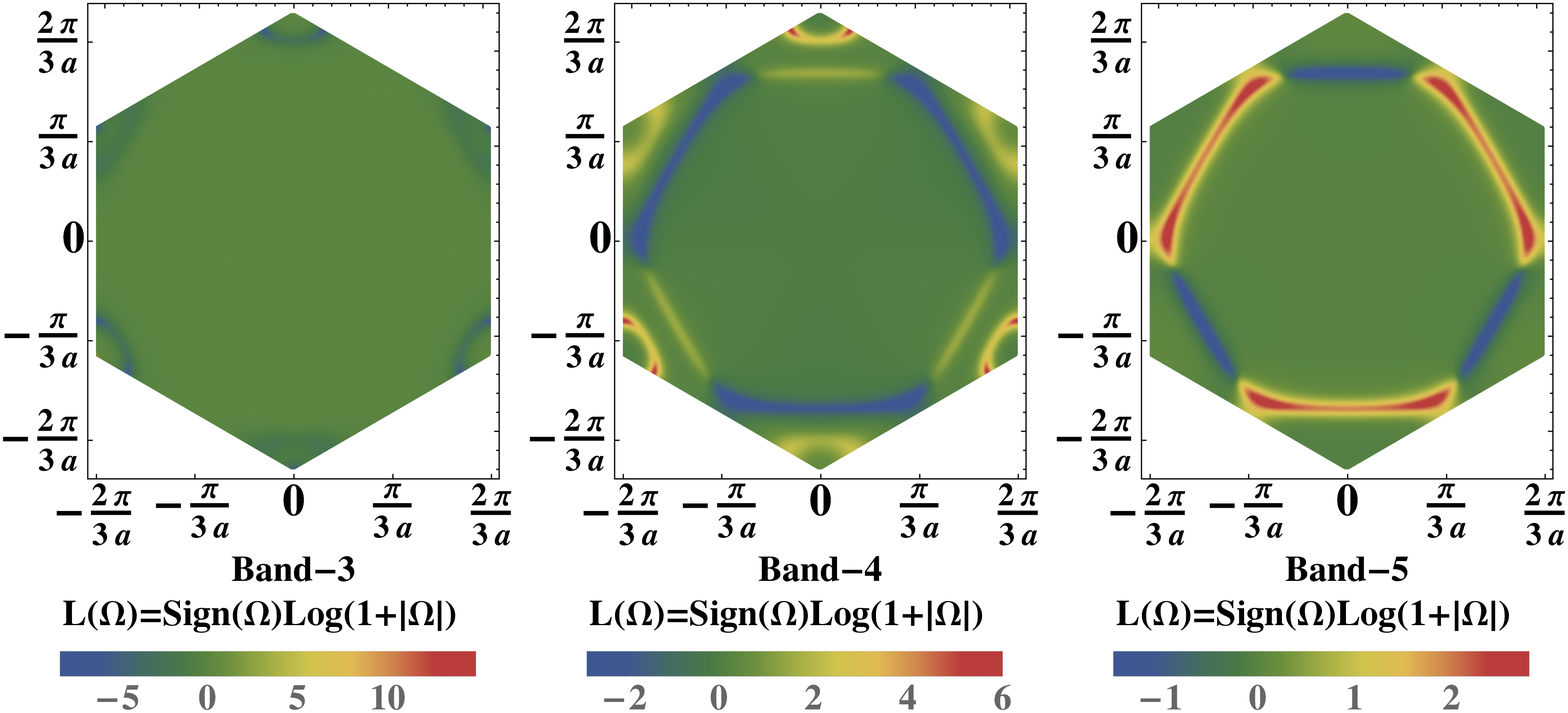}}
    \caption{(Color online.) Berry curvatures of the middle three anti-crossed bands in Fig.~\ref{fig:Dispersions}. Band numbers are ordered from bottom to top. When the magnetic field increases, there is one gapped ring around $\mathbf{\Gamma}$ between band 3 and 4 splits into two rings around $\mathbf{K}$ and $\mathbf{K}'$ leading to a topological phase transition.}
    \label{fig:BC}
\end{figure}

As shown in Fig.~\ref{fig:BC}, non-trivial Berry curvatures are induced around the anti-crossing regions, and thus the change of Chern numbers can be intuitively understood as a pair of gapped rings around $\mathbf{K}$ and $\mathbf{K}'$ combining into or split by one anti-crossing ring around $\mathbf{\Gamma}$. Notice that there are opposite Berry curvatures at $\mathbf{K}$ and $\mathbf{K}'$ in band-3 from the gap by spin-induced inversion symmetry breaking, but it does not contribute to the Chern number due to a cancellation between these two valleys \cite{zhang2015chiral}. However, as shown in Fig.~\ref{fig:CP_LH+}-\subref{fig:CP_HH-}, large phonon angular momentum $S^z_P=\left<\mathbf{u}^A_\mathbf{k}\times\mathbf{p}^A_{-\mathbf{k}}+\mathbf{u}^B_\mathbf{k}\times\mathbf{p}^B_{-\mathbf{k}}\right>$ occurs at $\mathbf{K}$ for band 2 and 3 giving rise to chiral phonons. The polarization of these phonons can be flipped by reversing the magnetic field and they can contribute to a valley Hall effect. 

Similar to the physics of gapped 2D Dirac systems \cite{bernevig2006quantum}, the physics of an anti-crossing magnon-phonon pair can be effectively described by
\begin{align}
    H_{\text{eff}}=\frac{E_{m\mathbf{k}}^{\pm}+E_{p\mathbf{k}}^{\pm}}{2}\mathcal{I}_{2}+\mathbf{d}^{\pm\pm}_\mathbf{k}\cdot\bm{\sigma}+V_\mathbf{k},
\end{align}
where $E_{m\mathbf{k}}^{+(-)}$ is the upper (lower) magnon energy without the DMI, $E_{p\mathbf{k}}^{+(-)}$ is the lower optical (upper acoustic) phonon energy without the DMI, $\bm{\sigma}=(\sigma_x, \sigma_y, \sigma_z)$ is the Pauli matrices, $\mathbf{d}^{\pm\pm}_\mathbf{k}$ opens a gap between $E_{m\mathbf{k}}^{\pm}$ and $E_{p\mathbf{k}}^{\pm}$ arising from the DMI and can be regarded as an analog of the gaping term in the Kane-Mele model \cite{kane2005quantum,kim2016realization}, and $V_\mathbf{k}$ includes terms that do not conserve particle numbers and perturbations that do not participate in opening the gap between the two bands \cite{go2019topological,zhang20203}. 
A Skyrmion (anti-Skyrmion) topological charge $Q\;(-Q)$ can then be defined with $\mathbf{d}^{\pm\pm}_\mathbf{k}$ as $Q=\frac{1}{4\pi}\int d^2\mathbf{k}\;\hat{\mathbf{d}}_\mathbf{k}\cdot\left(\partial_{k_x}\hat{\mathbf{d}}_\mathbf{k}\times\partial_{k_y}\hat{\mathbf{d}}_\mathbf{k}\right)$ for the upper (lower) band. In general, the analytical expression for $\mathbf{d}^{\pm\pm}_\mathbf{k}$ is not available, but since $\mathbf{d}^{\pm\pm}_z=(E_{m\mathbf{k}}^{\pm}-E_{p\mathbf{k}}^{\pm})/2$, the Skyrmion numbers will change with the moving of anti-crossing rings \cite{go2019topological}. As the band Chern number reflects the winding number of $\hat{\mathbf{d}}_\mathbf{k}$ wrapping the unit sphere in the Brillouin zone, a skyrmion arising from $\mathbf{d}$ with charge $Q$ determines the lower (upper) band with a Chern number $Q\;(-Q)$ \cite{bernevig2013topological}. In addition to changing the field strength, reversing the external field will also change the Chern numbers by flipping the sign, thus we find the topology of our system is highly tunable.

\paragraph{Thermal and Valley Hall effects.}In order to connect our results with possible experimental observations, we evaluate the thermal Hall effect rising from the non-trivial Berry curvature of magnon-polaron bands. With a longitudinal temperature gradient $\nabla_yT$, an anomalous transverse motion of magnon-polaron excitations can be induced by the fictitious field $\Omega_{n\mathbf{k}}^z$ associated with a transverse thermal conductivity $\kappa_{xy}$ as \cite{matsumoto2011rotational}
\begin{align}
    \kappa_{xy}=-\frac{k_B^2T}{\hbar V}\sum_{n,\mathbf{k}}\left[c_2(g(E_{n\mathbf{k}}))-\frac{\pi^2}{3}\right]\Omega_{n\mathbf{k}}^z,
\end{align}
where $c_2(x)=(1+x)\ln^2(1+1/x)-\ln^2x-2\text{Li}_2(-x)$, Li$_2(x)$  is the polylogarithm function, and $g(x)=(\text{exp}(x/k_B T)-1)^{-1}$ is the Bose-Einstein distribution.
\begin{figure}
\subfigure[Thermal Hall response. $\mathcal{B}$ is in unit of meV.]{\includegraphics[width=8.6 cm]{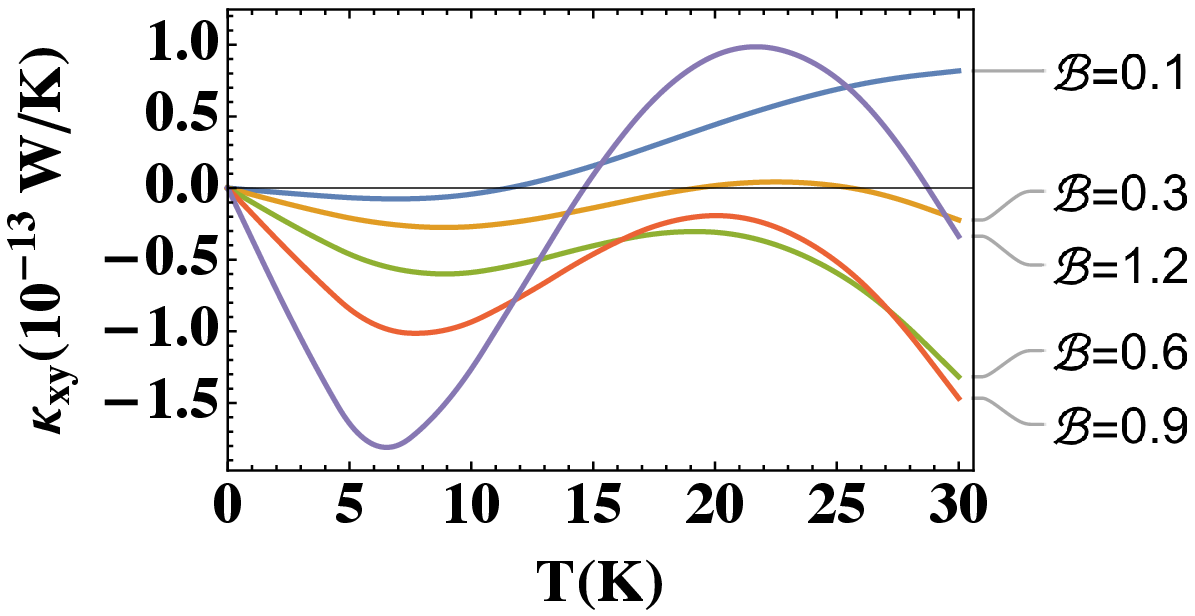}}
\subfigure[$\mathcal{B}=0.3$ meV with Chern number $(0,\;0,\;-2,\;+4,\;-2,\;0)$.]{\label{fig:CP1}\includegraphics[width=4.3 cm]{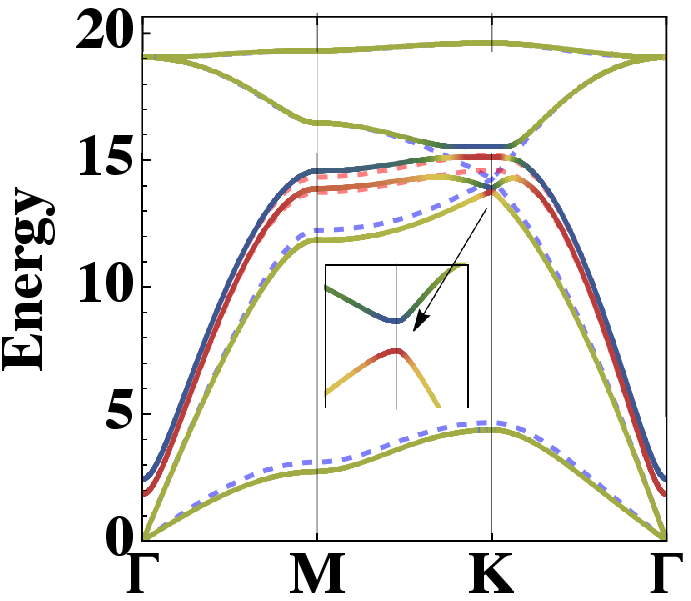}}\subfigure[$\mathcal{B}=1.2$ meV with Chern number $(0,\;+2,\;-2,\;+2,\;-2,\;0)$.]{\label{fig:CP2}\includegraphics[width=4.3 cm]{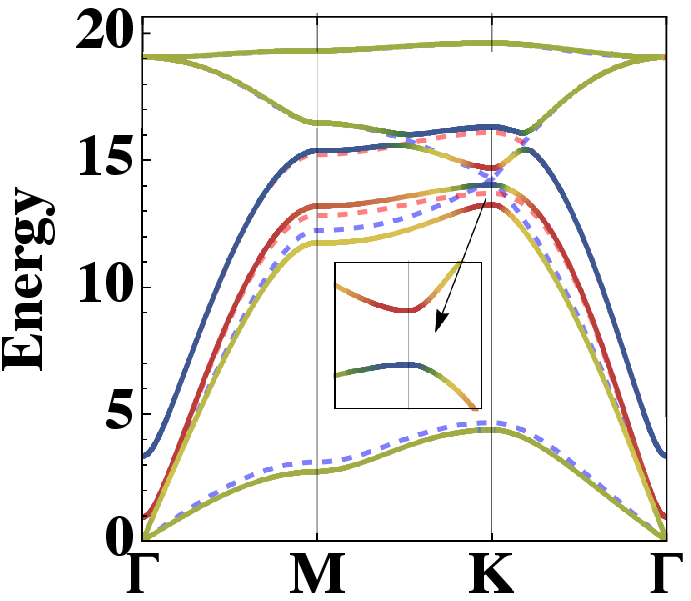}}
    \caption{(Color online.) (a) Thermal Hall response using parameters from MnPS$_3$. (b),(c) Band structures and Chern numbers for different external fields. See main text for details.}
    \label{fig:THC}
\end{figure}

In Fig.~\ref{fig:THC}, we evaluate $\kappa_{xy}$ with parameters \cite{wildes1998spin,joy1992magnetism,hashemi2017vibrational} for MnPS$_3$ as $m_A=m_B=M=55$ u, $S=5/2$, $J_1=1.54$ meV, $J_2=0.14$ meV, $g=-2.0$ and set $K_z=0.1$ meV, $D=0.5$ meV, $\hbar\sqrt{k_1/M}=11$ meV and $\hbar\sqrt{k_2/M}=2.2$ meV. At the low field, the two magnon bands couple with the lower optical phonon giving a Chern number distribution $(0,\;+2,\;-4,\;+2)$ from bottom to top, while they couple with the lower optical and upper acoustic phonon respectively at high field giving a Chern distribution $(-2,\;+2,\;-2,\;+2)$. These results are also consistent with our analysis on band topology by looking at the moving of gapped rings. The change of $\kappa_{xy}$ with magnetic field results from the topological transition with different Chern numbers, while the sign change with temperature reflects the competition among bands of different Chern numbers which come to dominate the transverse thermal transport.

In addition, as the spatial inversion symmetry is broken by the spin degree of freedom, the gap opens at $\mathbf{K}$ and $\mathbf{K}'$ valley, and thus gives rise to chiral phonons with different polarizations at these high symmetry points [see Fig.~\ref{fig:LH_Detail}\subref{fig:HH_Detail} and \ref{fig:CP1}\subref{fig:CP2}]. This has not been discussed in previous studies in coupled systems without optical phonons. By introducing a longitudinal strain gradient across the system, we expect opposite motion of chiral phonons at different valleys since $v\propto-\mathbf{E}_{\text{strain}}\times\mathbf{\Omega}$ in the transverse direction which creates a temperature difference between two edges \cite{zhang2015chiral}. As these two Hall effects originate from the non-trivial topology of the system, we expect to observe a thermal Hall signal only weakly affected by the bulk disorder.

\paragraph{Discussion.}In this Letter, we study the topology of magnon-polaron bands in a 2D honeycomb Neel order antiferromagnet with an in-plane DMI induced by magnon-phonon coupling. Without the DMI, the magnon or phonon bands are trivial, while non-trivial Berry curvature occurs around the anti-crossing rings opened by the magnon-phonon coupling. In contrast to previous studies, in our case, antiferromagnetic magnons can couple with both optical phonons and acoustic phonons giving rise to integer Chern numbers with an external magnetic field. Moreover, by changing the field magnitude and direction, it is possible to tune these Chern numbers along with changing the anti-crossing rings. 

We also investigated thermal Hall effects induced from finite Berry curvatures and propose valley Hall effects with chiral phonons arising from the inversion symmetry breaking by magnons. Even though we study the model on a honeycomb lattice, the coupling can be expressed with a displacement field $\mathbf{u}\approx\mathbf{u}_{ij}/a$ and a staggered spin field $\mathbf{n}\approx(\mathbf{S}_A-\mathbf{S}_B)/2S$ as $\frac{DS^2}{a^3}\left(\nabla\times \mathbf{u}\right)\cdot\left(\nabla\times\mathbf{n}\right)$ from Eq.~(\ref{Hmp}), which does not depend on lattice details \cite{sm}. 

This 2D model can also be generalized to a 3D system with mirror symmetry breaking in the bulk \cite{kim2019bulk,fernandez2019symmetry} and it can couple the magnons with out-of-plane phonon modes as well which could further enrich the physics of topology. In principle, our method can be used in any bosonic system such as plasmonics \cite{appelbaum2011proposal,di2013observation} and photonics \cite{ozawa2019topological}, and may find similar and interesting applications there. To our best knowledge, this is the first study on the topological properties arising from the coupling between antiferromagnetic magnons and optical phonons. Our work expands earlier studies where optical phonons are absent or ignored in the magnon-phonon coupling, and it may be useful to design tunable transport devices in the field of spintronics and draws a connection to chiral phonons with spin caloritronics.
\\
\paragraph{Acknowledge.}We thank Nemin Wei and Naichao Hu for helpful discussions on band topology. We gratefully acknowledge support from NSF DMR-1949701 and NSF DMR-2114825, with additional support from the NSF through the Center for Dynamics and Control of Materials: an NSF MRSEC under Cooperative Agreement No. DMR-1720595. This work was performed in part at the Aspen Center for Physics, which is supported by National Science Foundation grant PHY-1607611.


\bibliography{AFM_MP}
\end{document}


\preprint{APS/123-QED}

\title{Supplementary Materials for "Topological Magnon-Phonon in 2D AFM"}
\author{Bowen Ma}
\affiliation{
 Department of Physics, The University of Texas at Austin, Austin, Texas 78712, USA
}
\author{Gregory A. Fiete}
\affiliation{
 Department of Physics, Northeastern University, Boston, Massachusetts 02115, USA
}
\affiliation{Department of Physics, Massachusetts Institute of Technology, Cambridge, Massachusetts 02139, USA}

\date{\today}
             
\maketitle

\section{Lattice Vibration}
In this section, we derive the phonon Hamiltonian Eq.~(2) and the dynamical matrix $D(\mathbf{k})$ of Eq.~(5) in the main text.

We begin from the displacement length between site $i$ and $j$ as
\begin{align}
    |\Delta\mathbf{R}_{ij}|=|\mathbf{R}_{ij}|-|\mathbf{R}^0_{ij}|=\sqrt{(\mathbf{R}^0_{ij}+\mathbf{u}_{ij})^2}-|\mathbf{R}^0_{ij}|=|\mathbf{R}_{ij}^0|\sqrt{1+2\frac{\mathbf{R}_{ij}^0\cdot\mathbf{u}_{ij}}{|\mathbf{R}_{ij}^0|^2}+\mathbf{u}_{ij}^2}-|\mathbf{R}^0_{ij}|\approx\hat{\mathbf{R}}_{ij}^0\cdot\mathbf{u}_{ij}.
\end{align}
Therefore, to the lowest order, only in-plane displacement along bond $ij$ presents in the vibrational Hamiltonian, and by Hooke's law, we obtain Eq.~(2).

\begin{figure}
    \centering
    \includegraphics[height=6 cm]{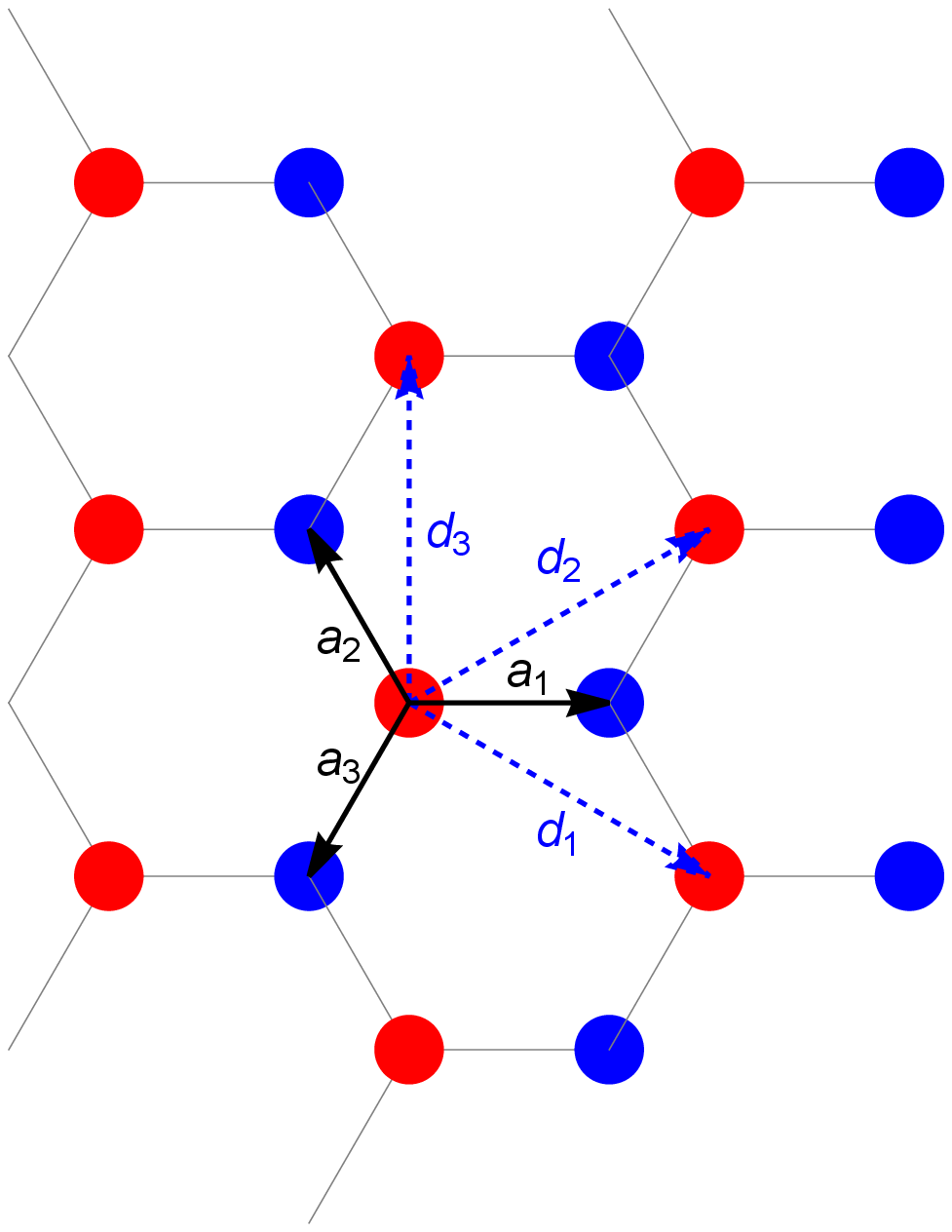}
    \caption{Honey comb lattice structure}
    \label{fig:S_lattice}
\end{figure}

To obtain $D(\mathbf{k})$, we perform a Fourier transform for $\mathbf{u}_i$ as $\mathbf{u}_i=\frac{1}{\sqrt{N}}\sum_\mathbf{k}u_{\mathbf{k}}e^{i\mathbf{k}\cdot\mathbf{R}_i}$ where $N$ is the number of unit cell, and take it into Eq.~(2) for honey comb lattice.
\begin{align}
    D(\mathbf{k})=\left[\begin{array}{cc}
    A(\mathbf{k})     & AB(\mathbf{k}) \\
    BA(\mathbf{k})     & B(\mathbf{k})
    \end{array}\right],
\end{align}
where
\begin{align}
     &A(\mathbf{k})= B(\mathbf{k})=\left[
\begin{array}{cc}
 \frac{3 k_1}{2}+3 k_2 \left(1-\cos{\frac{3 k_x a}{2}} \cos{\frac{\sqrt{3} k_y a}{2}}\right) & \sqrt{3} k_2 \sin{\frac{3 k_x a}{2}}\sin{\frac{\sqrt{3} k_y a}{2}} \\
 \sqrt{3} k_2 \sin{\frac{3 k_x a}{2}}\sin{\frac{\sqrt{3} k_y a}{2}} & \frac{3 k_1}{2}+k_2 \left(3-\cos {\frac{3 k_x a}{2}} \cos{\frac{\sqrt{3} k_y a}{2}}-2 \cos{\sqrt{3} k_y a}\right) \\
\end{array}
\right],\\
&BA(\mathbf{k})=AB^\dag(\mathbf{k})=\left[
\begin{array}{cc}
 -\frac{1}{2}k_1e^{\frac{i k_x a}{2} }\left(\cos{\frac{\sqrt{3}}{2} k_y a}+2e^{-i \frac{3k_x a}{2}}\right) & -i\frac{\sqrt{3}}{2} k_1 e^{\frac{i k_x a}{2}} \sin{\frac{\sqrt{3} k_y a}{2}} \\
 -i\frac{\sqrt{3}}{2}k_1 e^{\frac{i k_x a}{2}} \sin {\frac{\sqrt{3} k_y a}{2}} & -\frac{3}{2} k_1 e^{\frac{i k_x a}{2}} \cos{\frac{\sqrt{3} k_y a}{2}} \\
\end{array}
\right].
\end{align}
Here the basis is $\mathbf{u}_{\mathbf{k}}=\left(u^x_A(\mathbf{k}),u^y_A(\mathbf{k}),u^x_B(\mathbf{k}),u^y_B(\mathbf{k})\right)^T$ and the coordinates are shown in Fig.~\ref{fig:S_lattice} as $\mathbf{a}_1=a(1,0),\;
\mathbf{a}_2=a(-1/2, \sqrt{3}/2),\;\mathbf{a}_3=a(-1/2, -\sqrt{3}/2)$.
\section{Spin Hamiltonian}
In this section, we derive the magnon Hamiltonian $H_m(\mathbf{k})$ of Eq.~(5) in the main text by Holstein-Primakoff transformation, i.e.,
\begin{align}
\left\{\begin{array}{lll}
    S_{A,i}^+=S_{A,i}^x+iS_{A,i}^y=\sqrt{2S-a_i^\dag a_i}a_i\approx\sqrt{2S}a_i\\
    S_{A,i}^-=S_{A,i}^x-iS_{A,i}^y=a_i^\dag\sqrt{2S-a_i^\dag a_i}\approx\sqrt{2S}a_i^\dag\\
    S_{A,i}^z=S-a_i^\dag a_i
\end{array}\right.,\;
\left\{\begin{array}{lll}
    S_{B,j}^+=S_{B,j}^x+iS_{B,j}^y=b_j^\dag\sqrt{2S-b_j^\dag b_j}\approx\sqrt{2S}b_j^\dag\\
    S_{B,j}^-=S_{B,j}^x-iS_{B,j}^y=\sqrt{2S-b_j^\dag b_j}b_j\approx\sqrt{2S}b_j\\
    S_{B,j}^z=-S+b_j^\dag b_j
\end{array}\right..
\end{align}
Under this representation, Eq.~(1) can be re-written as
\begin{align}
    H_m&=J_1S\sum_{\left<ij\right>}(a_ib_j+a_i^\dag b_j^\dag+a_i^\dag a_i+b_j^\dag b_j)\nonumber\\
    &-J_2S\sum_{\left<\left<ii'\right>\right>}(a_i^\dag a_{i'}+a_{i'}^\dag a_i-a_i^\dag a_i-a_{i'}^\dag a_{i'})-J_2S\sum_{\left<\left<jj'\right>\right>}(b_j^\dag b_{j'}+b_{j'}^\dag b_{j}-b_j^\dag b_j-b_{j'}^\dag b_{j'})\nonumber\\
    &+\left(\mathcal{B}-\frac{K_z(1-2S)}{2}\right)\sum_ia_i^\dag a_i-\left(\mathcal{B}+\frac{K_z(1-2S)}{2}\right)\sum_jb_j^\dag b_j.
\end{align}
With Fourier transform $a_i\;(b_i)=\frac{1}{\sqrt{N}}\sum_\mathbf{k}a_i\;(b_i)e^{i\mathbf{k}\cdot\mathbf{R}_i}$, we obtain the expression for $\tilde{H}_m(\mathbf{k})$ in Eq.~(5) with basis $(a_{\mathbf{k}},b_{\mathbf{k}},a_{-\mathbf{k}}^\dag,b_{-\mathbf{k}}^\dag)^T$ as
\begin{align}
    \tilde{H}_m(\mathbf{k})=\left[\begin{array}{cc}
    A_m(\mathbf{k})&B_m(\mathbf{k})\\
        B_m^*(-\mathbf{k}) &A_m^*(-\mathbf{k})
    \end{array}\right],
\end{align}
where
\begin{align}
    &A_m(\mathbf{k})=S\left[\begin{array}{cc}
        3J_1+2J_2f(\mathbf{k})+K_z\left(1-\frac{1}{2S}\right)+\frac{\mathcal{B}}{S} & 0 \\
         0& 3J_1+2J_2f(\mathbf{k})+K_z\left(1-\frac{1}{2S}\right)-\frac{\mathcal{B}}{S}
    \end{array}
    \right],\;f(\mathbf{k})=3-\sum_{i}\cos{\left(\mathbf{k}\cdot\mathbf{d}_i\right)},\nonumber\\
    &B_m(\mathbf{k})=S\left[\begin{array}{cc}
        0 & J_1\sum_{i}e^{i\mathbf{k}\cdot\mathbf{a}_i} \\
         J_1\sum_{i}e^{-i\mathbf{k}\cdot\mathbf{a}_i}& 0
    \end{array}
    \right].
\end{align}
\section{Magnon-Phonon Coupling}
In this section, we derive $H_{mp}$ of Eq.~(4) from $H_D$ of Eq.~(3). As $\mathbf{D}_{ij}\propto\hat{\mathbf{z}}\times\mathbf{R}_{ij}$, we can determine DM vectors to be $\mathbf{D}_{ij}=D(\mathbf{R}_{ij})\hat{\mathbf{z}}\times\hat{\mathbf{R}}_{ij}$ ($D$ can be positive or negative for parallel or anti-parallel).Then the DMI between $A$-sublattice $i$ and $b$-sublattice $j$ can be written as
    \begin{align}
        \mathbf{D}_{ij}\cdot(\mathbf{S}_i\times\mathbf{S}_j)&=D(\mathbf{R}_{ij})(\hat{\mathbf{z}}\times\hat{\mathbf{R}}_{ij})\cdot(\mathbf{S}_i\times\mathbf{S}_j)=D(\mathbf{R}_{ij})\left[(\hat{\mathbf{z}}\cdot\mathbf{S}_i)(\hat{\mathbf{R}}_{ij}\cdot\mathbf{S}_j)-(\hat{\mathbf{z}}\cdot\mathbf{S}_j)(\hat{\mathbf{R}}_{ij}\cdot\mathbf{S}_i)\right]\nonumber\\
        &=SD(\mathbf{R}_{ij})\hat{\mathbf{R}}_{ij}\cdot(\mathbf{S}_i+\mathbf{S}_j)=SD(\mathbf{R}^0_{ij}+\mathbf{u}_{ij})(\hat{\mathbf{R}}^0_{ij}+\hat{\mathbf{u}}_{ji})\cdot(\mathbf{S}_i+\mathbf{S}_j),\label{S9}
    \end{align}
where we use relation $(\mathbf{A}\times\mathbf{B})\cdot(\mathbf{C}\times\mathbf{D})=(\mathbf{A}\cdot\mathbf{C})(\mathbf{B}\cdot\mathbf{D})-(\mathbf{A}\cdot\mathbf{D})(\mathbf{B}\cdot\mathbf{C})$ and $\hat{\mathbf{z}}\cdot\mathbf{S}_{i(j)}=\pm S$.

To the lowest order with in-plane $\mathbf{u}_{ij}$,
\begin{align}
    &D(\mathbf{R}^0_{ij}+\mathbf{u}_{ij})\approx D(\mathbf{R}^0_{ij})+\mathbf{u}_{ij}\cdot\nabla D=D(\mathbf{R}^0_{ij})+\frac{dD}{dR}\hat{\mathbf{R}}^0_{ij}\cdot\mathbf{u}_{ij}+\frac{dD}{Rd\theta}\hat{\theta}_{ij}\cdot\mathbf{u}_{ij},\label{S10}\\
    &\hat{\mathbf{R}}^0_{ij}+\hat{\mathbf{u}}_{ji}=\frac{\mathbf{R}^0_{ij}+\mathbf{u}_{ij}}{|\mathbf{R}^0_{ij}+\mathbf{u}_{ij}|}\approx\frac{\mathbf{R}^0_{ij}+\mathbf{u}_{ij}}{|\mathbf{R}^0_{ij}|}\left(1-\frac{\mathbf{R}^0_{ij}\cdot\mathbf{u}_{ij}}{|\mathbf{R}^0_{ij}|^2}\right).\label{S11}
\end{align}
Combining Eq.~(\ref{S10}) and (\ref{S11}) into Eq.~(\ref{S9}), we have 
\begin{align}
        \mathbf{D}_{ij}\cdot(\mathbf{S}_i\times\mathbf{S}_j)=SD(\mathbf{R}^0_{ij})\mathbf{R}^0_{ij}\cdot(\mathbf{S}_i+\mathbf{S}_j)+\frac{SD(\mathbf{R}^0_{ij})}{a}\left[\mathbf{u}_{ij}-(1+\gamma_1)\hat{\mathbf{R}}^0_{ij}\left(\hat{\mathbf{R}}^0_{ij}\cdot\mathbf{u}_{ij}\right)+\gamma_2\hat{\mathbf{R}}^0_{ij}\left(\hat{\mathbf{\theta}}_{ij}\cdot\mathbf{u}_{ij}\right)\right]\cdot(\mathbf{S}_i+\mathbf{S}_j),\label{S12}
    \end{align}
where $\gamma_1=-\frac{d\ln{D}}{d\ln{R}}$ and $\gamma_2=\frac{d}{d\theta}\ln{D}$.

The first term in Eq.~(\ref{S12}) is the zeroth order term that does not present in linear spin-wave Hamiltonian and  if we take $\gamma_1=\gamma_2=0$ which is reasonably small for tiny displacements, the second term reduce to Eq.~(4) in the main text. Moreover, if we begin from Eq,~(4) and take $a$ arbitrary small,
\begin{align}
    H_{mp}&=\frac{DS}{a}\sum_{\left<ij\right>}\left(\hat{\mathbf{R}}_{ij}^0\times \mathbf{u}_{ij}\right)\cdot\left[\hat{\mathbf{R}}_{ij}^0\times\left(\mathbf{S}_{A,i}+\mathbf{S}_{B,j}\right)\right]\nonumber\\
    &=\frac{DS}{2a}\sum_{ij}\left(\hat{\mathbf{R}}_{ij}^0\times \mathbf{u}_{ij}\right)\cdot\left[\hat{\mathbf{R}}_{ij}^0\times\left(\mathbf{S}_{A,i}+\mathbf{S}_{B,j}\right)\right]+\frac{DS}{2a}\sum_{ij}\left(\hat{\mathbf{R}}_{ji}^0\times \mathbf{u}_{ji}\right)\cdot\left[\hat{\mathbf{R}}_{ji}^0\times\left(\mathbf{S}_{A,j}+\mathbf{S}_{B,i}\right)\right]\nonumber\\
    &=DS^2\sum_{ij}\left(\hat{\mathbf{R}}_{ij}^0\times \frac{\mathbf{u}_{j}-\mathbf{u}_{i}}{a}\right)\cdot\left[\hat{\mathbf{R}}_{ij}^0\times\left(\frac{\mathbf{S}_{A,i}-\mathbf{S}_{B,i}}{2S}-\frac{\mathbf{S}_{A,j}-\mathbf{S}_{B,j}}{2S}\right)\right]\\
    &=-\frac{DS^2}{V}\int\;dV\left[\nabla\times \mathbf{u}(\mathbf{r})\right]\cdot\left[\nabla\times\mathbf{n}(\mathbf{r})\right],
\end{align}
where $\mathbf{u}(\mathbf{r})$ is the displacement field and $\mathbf{n}(\mathbf{r})=\frac{\mathbf{S}_A-\mathbf{S}_B}{2S}$ is the staggered spin field. It has similar form to Rashba spin-orbital coupling \cite{bychkov1984properties,manchon2015new} as $\alpha_R\mathbf{E}\cdot(\mathbf{p}\times\bm{\sigma})$ or Raman spin-phonon coupling \cite{zhang2010topological,sheng2006theory,kagan2008anomalous,wang2009phonon} as $\mathbf{h}\cdot(\mathbf{p}\times\mathbf{u})$.
\section{Bosonic Berry Curvature}
In this section, we introduce the numerical method for calculating bosonic Berry curvature and bosonic band Chern number.

We begin from a general bosonic BdG Hamiltonian for $N$ particles
\begin{align}
    H=\frac{1}{2}\sum_{\mathbf{k}}\mathbf{X}_{\mathbf{k}}^\dag \mathbf{H}_{\mathbf{k}}\mathbf{X}_{\mathbf{k}},\label{HX}
\end{align}
with a 2$N$ dimensional basis $\mathbf{X}_{\mathbf{k}}$ satisfying a commutator relation
\begin{align}
    [\mathbf{X}_{\mathbf{k}}, \mathbf{X}_{\mathbf{k}}^\dag]=g,\label{gX}
\end{align}
it can be transformed into a bosonic representation $\mathbf{Y}=Q^{-1}\mathbf{X}$ which satisfies \cite{del2004quantum}
\begin{align}
    H=\frac{1}{2}\sum_{\mathbf{k}}\mathbf{Y}_{\mathbf{k}}^\dag \mathbf{E}_{\mathbf{k}}\mathbf{Y}_{\mathbf{k}},\  [\mathbf{Y}_{\mathbf{k}}, \mathbf{Y}_{\mathbf{k}}^\dag]=\sigma_3,\ \mathbf{Y}_{\mathbf{k}}=\sigma_1\mathbf{Y}^*_{-\mathbf{k}}\sigma_1,\label{Y}
\end{align}
where $\sigma_3=\sigma_z\bigotimes\mathbb{I}_{N\times N},\;\sigma_1=\sigma_x\bigotimes\mathbb{I}_{N\times N}$ and $\mathbf{E}$ is a diagonal matrix.
By comparing Eq.~(\ref{HX})(\ref{gX}) with Eq.~(\ref{Y}), it can be seen that
\begin{align}
    g=Q[\mathbf{Y}, \mathbf{Y}^\dag]Q^\dag=Q\sigma_3Q^\dag\Rightarrow Q^\dag=\sigma_3Q^{-1}g,\;Q^\dag\mathbf{H}Q=\mathbf{E}\Rightarrow Q^{-1}g\mathbf{H}Q=\sigma_3\mathbf{E},\label{XY}
\end{align}
where the $n$-th column of $Q$ corresponds to the linear representation under basis $\mathbf{X}$ for eigenvector $\left|u_{n\mathbf{k}}\right>$ in the main text.
Then, with Eq.~(\ref{Y}) and (\ref{XY}),
\begin{align}
    i\hbar\frac{d}{dt}\mathbf{X}_{\mathbf{k}}=i\hbar Q_{\mathbf{k}}\frac{d}{dt}\mathbf{Y}_{\mathbf{k}}=Q_{\mathbf{k}}[\mathbf{Y}_{\mathbf{k}},\; H]=Q_{\mathbf{k}}[\mathbf{Y}_{\mathbf{k}},\; \frac{1}{2}\sum_{\mathbf{k}'}\mathbf{Y}_{\mathbf{k}'}^\dag \mathbf{E}_{\mathbf{k}'}\mathbf{Y}_{\mathbf{k}'}]=Q_{\mathbf{k}}\sigma_3\mathbf{E}_{\mathbf{k}}\mathbf{Y}_{\mathbf{k}}=Q_{\mathbf{k}}\sigma_3\mathbf{E}_{\mathbf{k}}Q_{\mathbf{k}}^{-1}\mathbf{X}_{\mathbf{k}}=g\mathbf{H}_{\mathbf{k}}\mathbf{X}_{\mathbf{k}}.
\end{align}
Therefore, a bosonic vector potential $\mathbf{A}_n$ and Berry curvature $\mathbf{\Omega}_{n\mathbf{k}}$ for $\mathbf{X}_{\mathbf{k}}$ can be defined as \cite{cheng2016spin,zhang2019thermal}
\begin{align}
    \mathbf{A}_n=i\left<u_{n\mathbf{k}}\right|g^{-1}\partial_{\mathbf{k}}\left|u_{n\mathbf{k}}\right>,\ \mathbf{\Omega}_{n\mathbf{k}}=i\left<\nabla_\mathbf{k}u_{n\mathbf{k}}\right|g^{-1}\times\left|\nabla_\mathbf{k}u_{n\mathbf{k}}\right>.\label{BC}
\end{align}
With the spirit of a momentum space lattice discretization method by Fukui {\textit{et al.}} \cite{fukui2005chern} and gauge-independent Berry curvature computation \cite{bernevig2013topological}, the Chern number and Berry curvature can be numerically calculated from Eq.~(\ref{BC}) once $Q_\mathbf{k}$ has been found.

\begin{figure}
    \centering
    \includegraphics[width=0.9\textwidth]{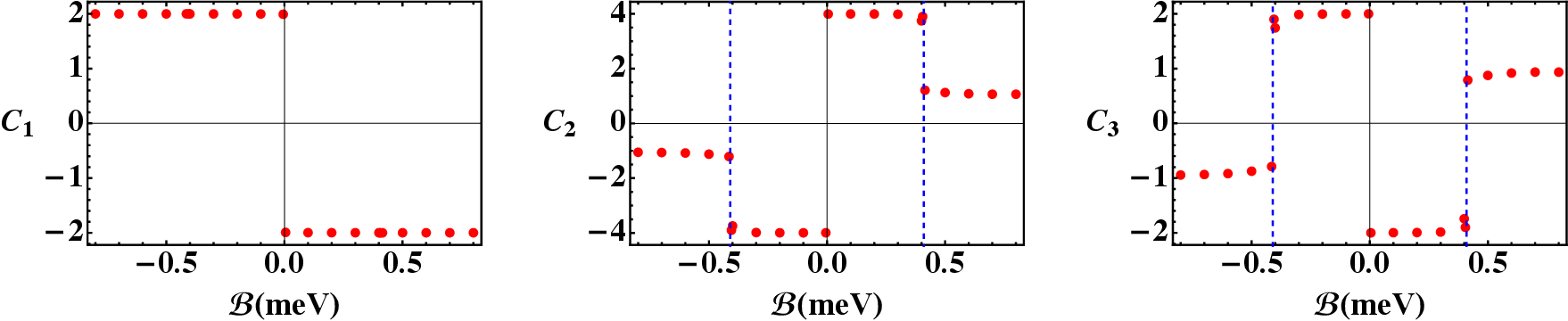}
    \caption{Chern numbers with varying the magnetic field. $C_1,C_2,C_3$ are the Chern number for the band from low to high. Blue dashed lines show the phase transition occur at $\mathcal{B}\approx\pm0.41$ meV.}
    \label{fig:CN}
\end{figure}

If there is no dispersionless Goldstone mode in the system, $\det{g\mathbf{H}}\neq0$ and the eigenvectors $P$ of $g\mathbf{H}$ can be found easily in any numerical methods, and if there is no degeneracy (or the degeneracy can be avoided in numerics), the eigenvectors can be rearranged so that
\begin{align}
    Q=PT,\label{Q=PT}
\end{align}
where $T=\text{Diag}(t_1,t_2,\dots,t_{2N})$.
Taking Eq.~(\ref{Q=PT}) into Eq.~(\ref{XY}), one can find
\begin{align}
    P^{-1}g(P^\dag)^{-1}=T\sigma_3T^\dag=\sigma_3 \text{Diag}(|t_1|^2,|t_2|^2,\dots,|t_{2N}|^2)\label{T},
\end{align}
which is a diagonal matrix and $|t_i|$ can be solved from Eq.~(\ref{T}).
Therefore, the paraunitary matrix $Q$ can be constructed from eigenvectors $P$ as
\begin{align}
    Q=P(P^\dag g^{-1}P\sigma_3)^{-\frac{1}{2}}U,
\end{align}
where $U$ is a $U(1)$ phase factor that can be chosen as the identity.
By this method, we numerically find the critical field $\mathcal{B}\approx0.41$ meV for the parameters used in main text [see Fig.~\ref{fig:CN}].
\section{Skyrmion vector}
In this section, we derive Eq.~(8) in the main text from Eq.~(5) by finding the quasiparticles without the DMI and then turning on the DMI as the interaction between quasiparticles.
For spin Hamiltonian $\tilde{H}_m$, since the ground state is collinear, it can be diagonalized by a Bogoliubov transformation $Q_m$ with a form
\begin{align}
    Q_m(\mathbf{k})=\left[\begin{array}{cccc}
    u_\mathbf{k}&0&0&-v_{-\mathbf{k}}  \\
    0&u^*_{\mathbf{k}}& -v^*_{-\mathbf{k}} &0\\
    0&-v^*_{\mathbf{k}} &u^*_{-\mathbf{k}} &0     \\
    -v_\mathbf{k}    & 0&0&u_{-\mathbf{k}}\\
    \end{array}\right],\;Q^\dag_m(\mathbf{k})\tilde{H}_m(\mathbf{k})Q_m(\mathbf{k})=\left[\begin{array}{cccc}
         E^+_{m\mathbf{k}}& & & \\
         & E^-_{m\mathbf{k}}& &\\
         & & E^+_{m,-\mathbf{k}}&\\
         & & & E^-_{m,-\mathbf{k}}
    \end{array}\right],
\end{align}
and it transforms the basis into Bogoliubov quasiparticle $(\alpha_{\mathbf{k}},\beta_{\mathbf{k}},\alpha^\dag_{-\mathbf{k}},\beta^\dag_{-\mathbf{k}})^T=Q_m^{-1}(\mathbf{k})(a_{\mathbf{k}},b_{\mathbf{k}},a^\dag_{-\mathbf{k}},b^\dag_{-\mathbf{k}})^T$.

For vibrational Hamiltonian, the dynamical matrix $D(\mathbf{k})$ is Hermitian and can then be diagonalized by a unitary matrix $Q_D(\mathbf{k})$ as
\begin{align}
    Q^\dag_D(\mathbf{k})\sqrt{M^{-1}}D(\mathbf{k})\sqrt{M^{-1}}Q_D(\mathbf{k})=\mathbf{E}^2_{p\mathbf{k}}=\left[\begin{array}{cccc}
       \left(E^t_{p\mathbf{k}}\right)^2& & & \\
         & \left(E^+_{p\mathbf{k}}\right)^2& &\\
         & & \left(E^-_{p\mathbf{k}}\right)^2&\\
         & & & \left(E^b_{p\mathbf{k}}\right)^2
    \end{array}\right],
\end{align}
where $M=\text{Diag}(m_A, m_A, m_B, m_B)$, $E^{t(b)}_{p\mathbf{k}}$ is the top(bottom) optical(acoustic) mode.
Then
\begin{align}
    &(\mathbf{u}_{-\mathbf{k}},\mathbf{p}_{\mathbf{k}})\left[
    \begin{array}{cc}
        \frac{1}{2}D(\mathbf{k})&0
        \\ 0&\frac{\mathcal{I}_{4}}{2M} 
    \end{array}\right]\left(\begin{array}{c}
        \mathbf{u}_{\mathbf{k}}  \\
         \mathbf{p}_{-\mathbf{k}} 
    \end{array}\right)\\
   & =\frac{1}{2}\left(
         \mathbf{u}_{-\mathbf{k}}\sqrt{M}Q_D(\mathbf{k})\sqrt{\mathbf{E}_{p\mathbf{k}}},\;
          \mathbf{p}_{\mathbf{k}}\sqrt{M^{-1}}Q_D(\mathbf{k})\sqrt{\mathbf{E}^{-1}_{p,-\mathbf{k}}}
    \right)\left[\begin{array}{cc}
         \mathbf{E}_{p\mathbf{k}}&  \\
         & \mathbf{E}_{p,-\mathbf{k}}
    \end{array}\right]\left(\begin{array}{c}
         \sqrt{\mathbf{E}_{p\mathbf{k}}}Q^\dag_D(\mathbf{k})\sqrt{M}\mathbf{u}_{\mathbf{k}}  \\
          \sqrt{\mathbf{E}^{-1}_{p,-\mathbf{k}}}Q^\dag_D(\mathbf{k})\sqrt{M^{-1}}\mathbf{p}_{-\mathbf{k}}
    \end{array}\right).
\end{align}
For a pure elastic vibration system, time reversal symmetry render $\mathbf{E}_{p\mathbf{k}}=\mathbf{E}_{p,-\mathbf{k}}$, and we have
\begin{align}
    Q_p(\mathbf{k})=\frac{1}{\sqrt2}\left[\begin{array}{cc}
        \mathcal{I}_4 & \mathcal{I}_4 \\
        -i\mathcal{I}_4 & i\mathcal{I}_4
    \end{array}\right]\left[\begin{array}{cc}
        \sqrt{M}Q_D(\mathbf{k})\sqrt{\mathbf{E}_{p\mathbf{k}}}&0  \\
          0&\sqrt{M^{-1}}Q_D(\mathbf{k})\sqrt{\mathbf{E}^{-1}_{p,-\mathbf{k}}}
    \end{array}\right],
\end{align}
which can properly diagonalize the phonon Hamiltonian without violating bosonic commutator with a quasiparticle basis ($\nu^t_{\mathbf{k}},\nu^+_{\mathbf{k}},\nu^-_{\mathbf{k}},\nu^b_{\mathbf{k}},\nu^{t\dag}_{-\mathbf{k}},\nu^{+\dag}_{-\mathbf{k}},\nu^{-\dag}_{-\mathbf{k}},\nu^{b\dag}_{-\mathbf{k}})^T=Q^{-1}_p(\mathbf{k})(\mathbf{u}_\mathbf{k},\mathbf{p}_{-\mathbf{k}})^T$.

Under quasiparticle representation $(\alpha_{\mathbf{k}},\beta_{\mathbf{k}},\nu^t_{\mathbf{k}},\alpha^\dag_{-\mathbf{k}},\beta^\dag_{-\mathbf{k}},\nu^+_{\mathbf{k}},\nu^-_{\mathbf{k}},\nu^b_{\mathbf{k}},\nu^{t\dag}_{-\mathbf{k}},\nu^{+\dag}_{-\mathbf{k}},\nu^{-\dag}_{-\mathbf{k}},\nu^{b\dag}_{-\mathbf{k}})^T$, the Hamiltonian transform to
\begin{align}
   H_\mathbf{k}=\frac{1}{2} \left[\begin{array}{cc}
         \begin{array}{cc}
            \mathbf{E}_{m\mathbf{k}}  &  \\
              & \mathbf{E}_{m,-\mathbf{k}}
         \end{array}& Q_m^\dag(\mathbf{k})(2\tilde{H}_{mp},0)Q_p(\mathbf{k}) \\
        Q_p^\dag(\mathbf{k})\left(\begin{array}{c}
             2\tilde{H}^\dag_{mp}  \\
             0
        \end{array}\right)Q_m(\mathbf{k}) & \begin{array}{cc}
        \mathbf{E}_{p,\mathbf{k}}      &  \\
              & \mathbf{E}_{p,-\mathbf{k}}
         \end{array} 
    \end{array}\right].
\end{align}
When ignoring particle number non-conserved terms and focusing on pair of bands that are anti-crossed, we only have two blocks for particles and holes respectively and it is enough to only look at particle space, which reads
\begin{align}
    H_{\text{eff}}=\left[\begin{array}{cc}
        E_{m\mathbf{k}}^{\pm} & C^\dag_{\mathbf{k}\pm\pm} \\
        C_{\mathbf{k}\pm\pm} & E_{p\mathbf{k}}^{\pm}
    \end{array}\right],
\end{align}
where
\begin{align}
\left[\begin{array}{cc}
    C_{\mathbf{k}++}  &C_{\mathbf{k}-+}  \\
     C_{\mathbf{k}+-}& C_{\mathbf{k}--}
\end{array}\right]=\left[\begin{array}{c}
    \sqrt{2m_AE^+_{p\mathbf{k}}}Q_{2}^\dag\tilde{H}^\dag_{mp} \\
    \sqrt{2m_BE^-_{p\mathbf{k}}}Q_{3}^\dag\tilde{H}^\dag_{mp}
\end{array}\right]\left[\begin{array}{cc}
    u_\mathbf{k}&0 \\
    0&u^*_{\mathbf{k}}\\
    0&-v^*_{\mathbf{k}}\\
    -v_\mathbf{k}    & 0\\
    \end{array}\right],\;Q_i\text{ is the i-th column of }Q_D.
\end{align}
Then $\mathbf{d}^{\pm\pm}_\mathbf{k}$ can be directly found from $C_{\mathbf{k}\pm\pm}$ as $\mathbf{d}^{\pm\pm}_\mathbf{k}=\left(\mathfrak{Re}\left[C_{\mathbf{k}\pm\pm}\right],\;\mathfrak{Im}\left[C_{\mathbf{k}\pm\pm}\right],\;\frac{ E_{m\mathbf{k}}^{\pm}- E_{p\mathbf{k}}^{\pm}}{2}\right)$. However, an analytical expression for $Q_D$ as well as $\mathbf{d}^{\pm\pm}_\mathbf{k}$ is in general not available.
\bibliography{sm}